\documentclass[conference]{IEEEtran}
\IEEEoverridecommandlockouts
\usepackage{cite}
\usepackage{amsmath,amssymb,amsfonts}
\usepackage{algorithmic}
\usepackage{multicol}
\usepackage{graphicx}
\usepackage{textcomp}
\usepackage{xcolor}
\usepackage{float}
\def\BibTeX{{\rm B\kern-.05em{\sc i\kern-.025em b}\kern-.08em
    T\kern-.1667em\lower.7ex\hbox{E}\kern-.125emX}}
\begin{document}
\newcommand{\norm}[1]{\left\lVert#1\right\rVert}

\title{Amplitude and Phase Estimation for Absolute Calibration of Massive MIMO Front-Ends\vspace{3pt}}
\author{\IEEEauthorblockN{Guoda Tian,
                Harsh Tataria,  and Fredrik Tufvesson}
\IEEEauthorblockA{Department of Electrical and Information Technology, Lund 
	University, Lund, Sweden}
e-mail: \{guoda.tian, harsh.tataria, fredrik.tufvesson\}@eit.lth.se\vspace{-9pt}}
\maketitle

%Abstract
\begin{abstract}
Massive multiple-input multiple-output (MIMO) promises significantly higher performance relative to conventional multiuser systems. However, the promised gains of massive MIMO systems rely heavily on the accuracy of the \emph{absolute front-end calibration}, as well as quality of \emph{channel estimates} at the base station (BS). In this paper, we analyze \emph{user equipment-aided} calibration mechanism to estimate the \emph{amplitude} scaling and \emph{phase drift} at each radio-frequency chain connected to the BS array. Assuming a uniform linear array at the BS and Ricean fading, we obtain the estimation parameters with \emph{moment-based} (amplitude, phase) and \emph{maximum-likelihood} (phase-only) estimation techniques. In stark contrast to previous works, we mathematically articulate the \emph{equivalence} of the two approaches for phase estimation. Furthermore, we rigorously derive a \emph{Cramér-Rao lower bound} to characterize the accuracy of the two estimators. Via numerical simulations, we evaluate the estimator performance with varying dominant line-of-sight powers, dominant angles-of-arrival, and signal-to-noise ratios. 
\end{abstract}

%Introduction 
\vspace{-6pt}
\section{Introduction}
\vspace{-3pt}
Fifth-generation (5G) systems are being deployed into commercial networks \cite{ERICSSON1}. The standardization efforts have resulted in a new radio access framework, known as Third Generation Partnership Project Release 15 (and beyond) \cite{3GPPTR38104}. A fundamental technology contributing to the spectral and energy efficiency targets of 5G systems is \emph{massive multiple-input multiple-output (MIMO)}. By scaling up the number of antennas at cellular base stations (BSs), massive MIMO sharply increases the \emph{beamforming gain} of the system, and enhances the ability to provide \emph{uniformly} good service to each user equipment (UE) \cite{MARZETTA1,LARSSON1,YANG1}. This has resulted in an order-of-magnitude \emph{increase} in the average spectral efficiency of 5G systems relative to their fourth-generation counterpart \cite{LARSSON1}. 

Since its inception in 2010, a vast amount of literature has developed around characterizing different performance aspects of massive MIMO systems (see e.g., \cite{SHAFI1,SHAFI2,MOLISCH1} for a summary). Nevertheless, the promised gains of massive MIMO greatly hinge on two key factors: (1) the  \emph{knowledge} of the channel state information at the BS and UE, (2) \emph{calibration quality} (precise definition presented later in the text). According to the related literature, massive MIMO calibration approaches are generally classified into two categories: namely, \emph{reciprocity} calibration \cite{R1,R2,R3,R4,R5} and \emph{absolute} calibration \cite{A1,A2,A3,A4,A5,A6,Intra}. Reciprocity calibration is required in massive MIMO to ensure that the downlink channel is \emph{reciprocal} to the uplink. The concept of the relative reciprocity calibration was first introduced in \cite{R1}. Extending this, a high-level network protocol of UE synchronization and reciprocity-based calibration was presented in \cite{R2}. Moreover, the authors of \cite{R3,R4} derived several practical approaches for reciprocity calibration and validated the results in real-time via the Lund University massive MIMO testbed. A taxonomy of the existing reciprocity calibration methods with an antenna grouping strategy is proposed in \cite{R5} to \emph{shorten} the calibration time. In  contrast to reciprocity calibration, absolute calibration, is required for angle-of-arrival (AOA) estimation and positioning. Absolute calibration exploits the amplitude and phase spectra \emph{across} the BS array, as shown in \cite{A1,A2}. Approaches such as \emph{intra-array} and \emph{UE-aided} calibration are discussed in \cite{A3,A4,A5}. The authors of \cite{A6} combine array calibration with AOA estimation, while the authors of \cite{Intra} propose mutual coupling-based methods for estimating the phase and amplitude relationships between each radio-frequency (RF) chain at the BS. 
 
The intra-array based calibration can be implemented either with or without transmission lines between antenna elements. The later case outperforms the former in terms of \emph{interconnect flexibility} at the cost of calibration accuracy, since its performance \emph{degrades} with increasing electrical distance between successive antennas \cite{A3}. For UE-aided calibration, a better trade-off between the flexibility and accuracy is expected, and is therefore worth further investigation. To our best knowledge, prior works on UE-aided calibration only consider simple additive white Gaussian noise (AWGN) channels, which naturally do not reflect the physics of wave propagation. To this end, we analyze UE-aided absolute calibration over a Ricean fading channel, often used to model dominant line-of-sight (LOS) components in addition to diffuse multipath components \cite{MOLISCH1,TSE1}. We provide a methodology to analyze two types of practical estimators (described later in the text) and derive the corresponding Cramér-Rao lower bound (CRLB) for evaluating the quality of amplitude and phase estimates. 

Our main contributions are as follows: For an uplink single-user massive MIMO system, assuming a uniform linear array (ULA) at the BS and Ricean fading propagation, we establish two general, yet practical, analytical approaches to estimate the amplitude scaling and phase drift associated with each RF chain. The first approach is based on \emph{moment-based} estimation of the aforementioned parameters, while the second is based on \emph{maximum-likelihood} estimation (MLE), for obtaining phase estimates. We mathematically show that both estimators have an equivalent form when estimating  the phase of the RF chains, and back up the mathematical findings with the required physical intuition.  For evaluating the accuracy of both estimators, we derive the CRLB to characterize the fundamental lower limits on error of the estimated phase and amplitude scaling coefficients across the array. \emph{To the best of our knowledge, this has been missing from the literature.} We evaluate the derived estimator performance on a ULA-based \emph{numerical} framework. We show that under the presence of dominant LOS conditions, the \emph{variance} of the phase estimates rapidly converges to the predicted CRLB for both estimator types. In addition, the amplitude and phase estimation accuracies of both approaches significantly improve with growing LOS powers and signal-to-noise ratios (SNRs).

%System Model
\begin{center}
  \begin{figure}[t!]
  	\hspace{15pt}
  	\includegraphics[width=0.84\linewidth]{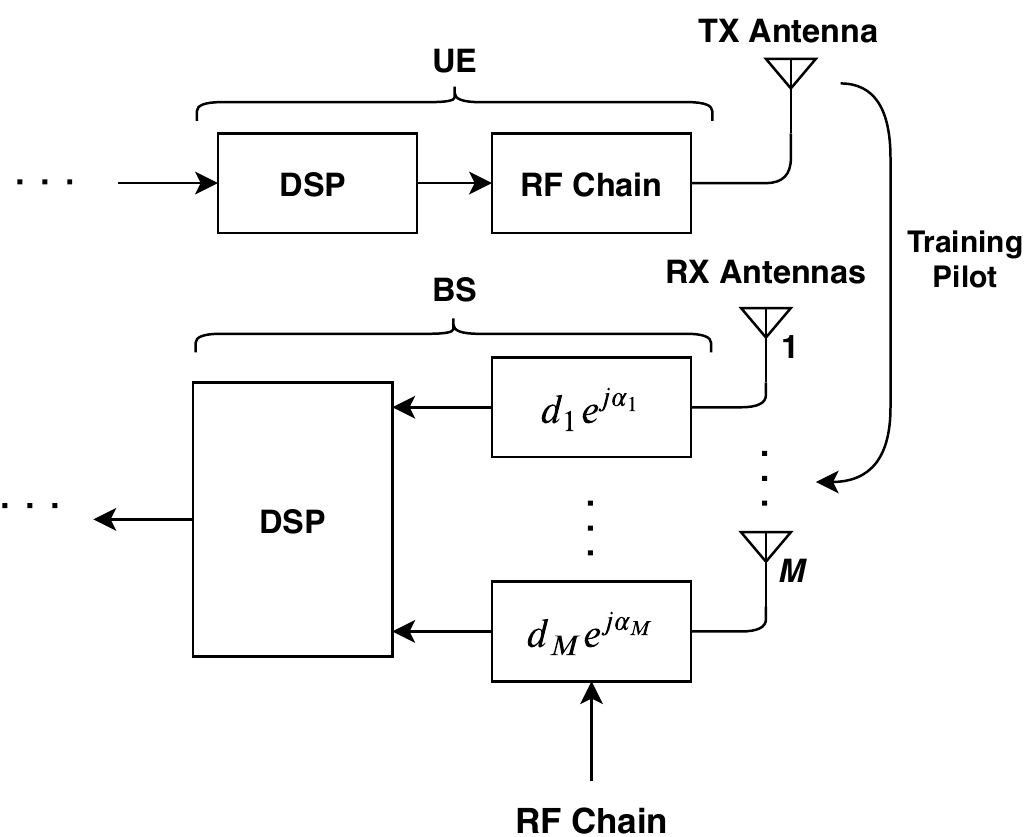}
  	\vspace{-7pt}
  	\caption{A single-user uplink massive MIMO system with pilot transmission from the UE to the $M$ BS antennas, which are interfaced with $M$ RF chains.}
  	\label{SystemModel} 
  	\vspace{-11pt}
  \end{figure}
\end{center}
\vspace{-27pt}
\section{System Model}
\label{SystemModelS}
We consider the \emph{uplink} of a single-user massive MIMO system, which has $M$ antenna elements configured in a ULA at the BS. We assume reciprocity-based operation in the time-division duplex mode where the UE sends uplink pilot signals, which are used to \emph{estimate} the calibration parameters at the $M$ RF chains interfacing with the receive antennas. The overall system model is depicted in Fig.~\ref{SystemModel}. We assume narrowband propagation between the UE and the BS, with uniform power allocation. More specifically, we employ the use of a general \emph{Ricean} fading model, where the small-scale fading impulse response is an amalgamation of a dominant LOS component, in addition to the diffuse multipath components. The LOS component is governed by the far-field array steering vector in a given direction, and the diffuse components are modeled as complex Gaussian random variables (exact definition later in the text). The use of such model is rather popular in massive MIMO performance evaluation, particularly in urban scenarios where many diffuse paths are expected with some dominant LOS components \cite{TATARIA1,TATARIA2,WEISEMANN1}.  Considering this, the received signal observation vector, $\mathbf{y}_t \in \mathbb{C}^{M \times 1}$ during time $t$ can be written as
\vspace{-7pt}
\begin{equation}
\label{signalmodel}
	\mathbf{y}_t = \underbrace{\gamma_t \hspace{1pt}\mathbf{D}_t \hspace{1pt}\mathbf{a}(\phi_t)\hspace{1pt}p_t}_{\mathbf{s}_t} + \underbrace{\mathbf{D}_t\mathbf{h}_t\hspace{1pt}p_t+\mathbf{n}_t}_{\boldsymbol{\omega}_t},
	\vspace{-1pt}
\end{equation}
 where $\gamma_t$ and $p_t$ are scalar quantities which denote the large-scale LOS power and the pilot transmitted by the UE at time $t$. The power contained in $p_t$ is normalized to unity, such that $|p_t|^2=1$, over all values of $t=1,2,\dots,T$. Since we consider a ULA, the array steering vector is a known function of the azimuth AOA, which is denoted as $\mathbf{a}(\phi_t)\in \mathbb{C}^{M\times 1}$ with an incoming angle $\phi_t$. In addition, the vectors $\mathbf{h}_t\in \mathbb{C}^{M\times 1}$ and $\mathbf{n}_t\in \mathbb{C}^{M\times 1}$ denote the diffuse multipath components and the AWGN at time $t$, such that 
 $\mathbf{h}_t\sim{}\mathcal{CN}(0,\sigma^2)$ and $\mathbf{n}_t\sim\mathcal{CN}(0,N_0/2)$. To this end, the mean of the diffuse components is zero and the variance (power) is $\sigma^2$ across all $t=1,2,\dots,T$. Likewise, the mean of the AWGN at the BS is zero and variance is $N_0/2$. Following this, the SNR at time $t$ is given by $|\hspace{1pt}p_t|^2/(N_0/2)$. The diagonal matrix, $\mathbf{D}_t\in \mathbb{C}^{M\times M}$, contains the $M$ \emph{amplitude scaling} and the \emph{phase drift} entries for each RF chain. This matrix models the random phase and amplitude changes introduced by phase jitter at the local oscillators, and RF signal conditioning units such as low-noise amplifiers and active bandpass filters. We note that $\mathbf{D}_t = \textrm{diag}(d_1e^{j\alpha_1},d_2e^{j\alpha_2},...,d_Me^{j\alpha_M})$. We further assume that $\mathbf{h}_t$ and $\mathbf{n}_t$ are statistically independent and  $\mathbf{n}_t$ is uncorrelated over $t=1,2,\dots,T$. With the above in mind, the auto-correlation at time $t$, $\mathbb{E}\{\boldsymbol{\omega}_{\hspace{-1pt}t}\hspace{1pt}
 \boldsymbol{\omega}^{H}_{\hspace{-1pt}t}\}$, can be evaluated as 
 \vspace{-2pt}
\begin{align}
    \nonumber
    \mathbb{E}\left\{\boldsymbol{\omega}_t\boldsymbol{\omega}^H_t\right\}&= \mathbb{E}\left\{(\hspace{1pt}\mathbf{D}_t\mathbf{h}_t\hspace{1pt}p_t+\mathbf{n}_t)
    (\mathbf{D}_t\mathbf{h}_t\hspace{1pt}p_t+\mathbf{n}_t)^H\right\}\\ \label{autocorr}
    & = N_0\hspace{1pt}\mathbf{I}_M + \mathbf{D}_t\hspace{1pt}|p_t|^2\sigma^2
    \hspace{1pt}\mathbf{D}_{t}^H, \\[-18pt]\nonumber
\end{align}
where $\mathbf{I}_M$ denotes the $M\times{}M$ identity matrix. Moreover, by definition, the cross-correlation between two time intervals, namely $t=1$ and $t=2$, can be expressed as
\vspace{-3pt}
 \begin{equation}
 \label{crosscorr}
     \mathbb{E}\hspace{1pt}\{\boldsymbol{\omega}_{t=1}\boldsymbol{\omega}_{t=2}^{H}\} = \sigma^2\hspace{1pt}\mathbf{D}_t\hspace{1pt}\mathbf{D}_t^H.
     \vspace{-3pt}
 \end{equation}
 Between multiple time instances, the channel, $\mathbf{h}_t$, is assumed to be changing in accordance with its definition. This can be caused by small changes in the UE position, or mobility of objects in the propagation environment. For simplicity, from here onward, we drop the subscript $t$ used in the right-hand side of \eqref{signalmodel}, and assume that all further computations are performed at a given time instance $t$. Therefore, the received vector $\mathbf{y}_{\hspace{-1pt}t}$ follows a complex Gaussian distribution given by
 \vspace{-1pt}
 \begin{equation}
    \label{totalyt}
 	\mathbf{y}_{\hspace{-1pt}t} \sim \mathcal{CN} \left(\gamma\hspace{1pt}\mathbf{D}
 	\hspace{1pt}\mathbf{a(\phi)}\hspace{1pt}p,  N_0\hspace{1pt}\mathbf{I}_M + \mathbf{D}\hspace{1pt}|p|^2\sigma^2\hspace{1pt}\mathbf{D}^H\right).
 	\vspace{-1pt}
 \end{equation}  
  Given the model in \eqref{signalmodel}-\eqref{totalyt}, $\mathbf{y}_t$, $p$ and $\mathbf{a}(\phi)$ are assumed to be known by the BS. Other parameters such as $\gamma$, $\sigma^2$, $\mathbf{D}$, $\mathbf{h}$, $\mathbf{n}$ are assumed to be \emph{unknown}, which is the case in practice. Observing over $T$ intervals, the \emph{composite} received signal is given by stacking all $\mathbf{y}_t$ across  $t=1,2,\dots{},T$ obtaining
  \vspace{-1pt}
 \begin{equation}
     \label{totaly}
 	\mathbf{y} \sim \mathcal{CN}\big(\underbrace{\mathbf{1}\hspace{-1pt}\otimes\hspace{-1pt}\gamma\hspace{1pt} p\hspace{1pt}\mathbf{D}\mathbf{a(\phi)}}_{\boldsymbol{\mu}(\boldsymbol{\xi})}, \hspace{3pt}\underbrace{\mathbf{I}_T\hspace{-1pt}\otimes\hspace{-1pt} \mathbf{I}_{M}N_{0} \hspace{-1pt}+\hspace{-1pt}\widetilde{\mathbf{I}}_T\hspace{-1pt}\otimes
 	\hspace{-1pt}\sigma^2\hspace{1pt}\mathbf{D}\mathbf{D}^H}_{\mathbf{C}\hspace{1pt}(\boldsymbol{\xi})}
 	\big),
 	\vspace{-2pt}
 \end{equation}
 where $\mathbf{1}\in \mathbb{R}^{T\times1}$ is a column vector of unit entries, $[1,1,\dots,1]^T$, while $\mathbf{\widetilde{I}}_T = \boldsymbol{1}.(\boldsymbol{1}^T) \in \mathbb{R}^{T\times T}$ is a matrix containing unit entries. In addition, $\otimes$ is the Kronecker product operation, $\boldsymbol{\mu}(\boldsymbol{\xi})$ denotes the mean vector and $\mathbf{C}(\boldsymbol{\xi})$ denotes the variance over all $T$ time intervals. Note that the vector argument $\boldsymbol{\boldsymbol{\xi}}$ contains the unknown quantities in $d,\alpha,\sigma^2$ and $\gamma$, respectively. That is, $\boldsymbol{\xi}=[d_1,d_2,\dots,d_M,\alpha_1,...,\alpha_M,\sigma^2,\gamma]^T$. With this setup, the subsequent section of the paper discusses the phase and amplitude estimation techniques with the aim to calibrate $M$ RF chains at $M$ antennas of the BS.

%Estimator Design
\section{Estimator Design}
\label{EstimatorDesign}
\vspace{-2pt}
In order to perform absolute calibration, one needs to estimate the subspace spanned by the vector $\mathbf{v} = \{d_1e^{j\alpha_1},d_2e^{j\alpha_2},...,d_Me^{j\alpha_M}\}$. According to this requirement, we analyze \emph{two} estimators, namely the moment-based estimator and the MLE estimator for estimating the phase drift vector $\boldsymbol{\alpha} = [\alpha_1, \alpha_2, ..., \alpha_M]^T$ and a moment-based estimator for estimating the vector space $\mathbb{R}^{M \times 1}$ spanned by the magnitude vector $\mathbf{d} = [d_1, d_2,...,d_M]^T$ of the RF chains. Later on, we prove that the analytical expression of the moment-based phase drifting estimator coincides with the MLE-based estimator. 
\vspace{-2pt}
\subsection{Phase Estimation}
\subsubsection{Moment-Based Estimator}
\vspace{-1pt}
From \eqref{totaly}, we can see that the information regarding the RF phase drifts is only embedded in the \emph{first-order} statistics of the composite received signal. Enlightened by this, we analyze the moment-based estimator which computes the \emph{expectation} of the composite vector $\mathbf{y}$ before estimating the phase drifts. The phase vector $\boldsymbol{\alpha} = \big[\alpha_1, \alpha_2,\dots,\alpha_M]^T$ is estimated by
 \begin{equation}
 \label{momentbasedestresult}
 	\boldsymbol{\hat{\alpha}} = \arg\left\{\sum_{t=1}^{T}\mathbf{y}_t\right\} - \arg\left\{\mathbf{a}(\phi)\right\} -\arg\left\{\mathbf{1}p\right\}.
 	\vspace{2pt}
 \end{equation}
\emph{Sketch of Proof:} Via some straightforward algebra, one can show that  $\mathbb{E}\{\mathbf{y}_t\} = \gamma{}\hspace{1pt}\mathbf{D}\mathbf{a}(\phi)p$. 
With a \emph{large} total observation time in $T$, one can expect that the empirical probability distribution of $\mathbf{y}_t$ converges almost surely to its true probability distribution. Using this fact allows us to accurately approximate the 
moments (first-order only, since second-order contains no utilizable information) of the 
empirical distribution with the true distribution, such that $\frac{1}{T}\sum_{t=1}^{T}\mathbf{y}_t\approx\hspace{-1pt}\mathbb{E}\{\mathbf{y}_t\}$. Taking $\arg\{\frac{1}{T}\sum_{t=1}^{T}\mathbf{y}_t\}=\arg\{\mathbf{a}(\phi)\}+\arg\{\mathbf{1}p\}+\boldsymbol{\alpha}$, and solving for $\boldsymbol{\alpha}$ yields the desired phase estimate. \vspace{1pt}
\subsubsection{MLE}
If $\mathbf{y}$ has the probability distribution function $\bar{p}(\boldsymbol{\xi},\boldsymbol{y})$, then the MLE formulates an optimization problem on the maximization of the log-likelihood function. That is
\begin{align}
\nonumber
\boldsymbol{\hat{\alpha}} &= \arg \max \left\{\bar{p}\hspace{1pt}(\boldsymbol{\xi},\mathbf{y})\right\} \\[2pt] 
\label{MLEProb}
&\overset{(a)}= \min_{\mathbf{\alpha}}\left\{\ln\det\left(\mathbf{\Lambda}\right)+\hspace{1pt} \boldsymbol{\beta}^H \mathbf{C}^{-1}\left(\boldsymbol{\xi}\right)\boldsymbol{\beta}\right\},\\[-16pt]&\nonumber \end{align}
where $\boldsymbol{\beta} = \mathbf{y}-\left(\mathbf{1}\otimes\gamma\hspace{1pt}\mathbf{D}\hspace{1pt}\mathbf{a}(\phi)\right)$, $\mathbf{C}\left(\boldsymbol{\xi}\right)=\tilde{\mathbf{Q}}^H\mathbf{\Lambda} \tilde{\mathbf{Q}}$, with $\tilde{\mathbf{Q}}=\mathbf{Q}\otimes \mathbf{I}_{M}$ and $\mathbf{Q}\in \mathbb{C}^{T\times T}$ is defined as the normalized discrete Fourier transform (DFT) matrix. Moreover,  $\mathbf{\Lambda}$ is defined as
\vspace{2pt}
\begin{equation*}
	\mathbf{\Lambda} = 
	\begin{bmatrix}
	\sigma^2T\mathbf{D}\mathbf{D}^H & & &\\
	 & \ddots& & \\
	& & & \boldsymbol{0}
	\end{bmatrix} + N_0\hspace{1pt}\mathbf{I}_{MT},
	\vspace{1pt}
\end{equation*}
and is a $MT \times MT$ matrix. In \eqref{MLEProb}, $(a)$ is a result of equivalently minimizing the argument of the exponential function in $\bar{p}(\boldsymbol{\xi},\boldsymbol{y})$.
Substituting $\boldsymbol{1} \otimes \mathbf{D} \mathbf{a(\phi)} = \sqrt{T}\tilde{\mathbf{Q}}^H\begin{bmatrix}
\mathbf{I} \dots \mathbf{0}
\end{bmatrix}^T\mathbf{Da(\phi)}$ in \eqref{MLEProb} and simplifying yields 
\vspace{-3pt}
   \begin{align}
   \nonumber
   \boldsymbol{\hat{\alpha}} &= \mathbf{\min_{\alpha}}\{\ln \det\mathbf{\Lambda} +  \mathbf{y}^H\mathbf{C}^{-1}(\boldsymbol{\xi})\hspace{1pt}\mathbf{y} \\ \nonumber
   &- 2\gamma\sqrt{T}\Re\Big[\mathbf{a}^H\hspace{-1pt}(\phi)\hspace{1pt}
   \mathbf{D}^H\big[(\sigma^2T\mathbf{DD}^H\hspace{-2pt}+\hspace{-3pt}
   N_0\hspace{1pt}\mathbf{I}_M)^{-1}
   \hdots \mathbf{0} \big]\tilde{\mathbf{Q}}\hspace{1pt}\mathbf{y}\Big] \\ \label{MLESimp}
   &+\gamma^2T\mathbf{a}^H\hspace{-1pt}(\phi)\mathbf{D}^H(\sigma^2T\mathbf{DD^H}+N_0\mathbf{I}_M)^{-1}\mathbf{D\hspace{1pt}a(\phi)}\}.
    \end{align}
Note that $\Re[\cdot]$ denotes the real component of a complex quantity. From \eqref{MLESimp}, it is clear that the phase information is only contained in the term $\Re\hspace{1pt}[\mathbf{a}^H\hspace{-1pt}(\phi)\hspace{2pt}\mathbf{D}^H[\hspace{1pt}(\sigma^2\hspace{1pt}T\hspace{1pt}\mathbf{DD}^H\hspace{-2pt}+\hspace{-2pt}N_0\mathbf{I}_M)^{-1}\hdots \boldsymbol{0}\hspace{1pt}]\mathbf{\tilde{Q}\hspace{1pt}y}]$. Thus, can derive the MLE of $\boldsymbol{\alpha}$ as 
\vspace{2pt}
\begin{align}
\nonumber
\label{Esteq}
\hat{\boldsymbol{\alpha}} &\hspace{-1pt}=\hspace{-1pt}\max_{\alpha}\hspace{-1pt}\left\{ \Re\Big[\mathbf{a}^H\hspace{-1pt}(\phi)\hspace{1pt}\mathbf{D}^H\big[\hspace{1pt}(\sigma^2{}T\hspace{1pt}\mathbf{DD}^H\hspace{-3pt}+\hspace{-3pt}N_0\hspace{1pt} \mathbf{I}_M)^{-1}
\hspace{-3pt}\hdots \boldsymbol{0} \big]\tilde{\mathbf{Q}}\hspace{1pt}\mathbf{y}\Big]\right\} \nonumber \\ 
& = \arg\left\{\sum_{t=1}^{T}\mathbf{y}_t\right\} - \arg\left\{\mathbf{a}(\phi)\right\} -\arg\left\{\mathbf{1}p\right\}. \\[-17pt]
&\nonumber
\end{align}
\emph{The above result is mathematically equivalent to the one derived from the moment-based estimator in \eqref{momentbasedestresult}.} The intuition behind this equivalence can be explained as follows: 
The received vector $\mathbf{y}$ follows complex Gaussian distribution, and hence the first and second-order statistics of $\mathbf{y}$ contain the vast majority of its underlaying information. To this end, the \emph{optimal} solution can be found by exploiting the first and second-order statistics \cite{ICA}. For the moment-based estimator, since the second-order statistics do not contribute to the phase estimates, the first-order statistics can be used to derive an optimal estimator, which is identical to the MLE. 
\vspace{-3pt}
\subsection{Amplitude Estimation}
\vspace{-2pt}
 We now analyze the moment-based estimator for deriving the amplitude scaling coefficients of the $M$ RF chains. We refrain from utilizing the MLE for amplitude estimation as the presence of higher-order terms makes maximization of the log-likelihood function a mathematically complex task. We estimate the vector space spanned by $\mathbf{d}$. Unlike for phase estimation, since \emph{both} the first and second-order statistics of $\mathbf{y}$ contains useful information, it is necessary to estimate the covariance matrix $\mathbf{C(\boldsymbol{\xi})}$, which we denote as 
 $\hat{\mathbf{C}}(\boldsymbol{\xi})$. We observe that the upper and lower triangular block diagonal sub matrices of $\hat{\mathbf{C}}(\boldsymbol{\xi})$ contain the relevant terms for $\sigma^2\mathbf{d}\odot\mathbf{d}$, which can be extracted for estimation. Note that $\odot$ denotes the Hadamard product. We therefore provide a \emph{closed-from} solution for the moment-based amplitude estimator as
 \begin{equation}
 \label{mom}
 \hat{\mathbf{d}}=\sqrt{\sum\limits_{t=1}^{T}\mathbf{y}_t\odot\mathbf{y}_t^{*}+
 \textrm{vecdiag}\bigg[\sum\limits_{t=1}^{T}\sum\limits_{\substack{t'=1\\t'\neq{}t}}^{T}
 \tilde{\mathbf{C}}\left(\boldsymbol{\xi}\right)\hspace{-1pt}|_{\left(t,t'\right)}\bigg]},
 \vspace{-2pt}
 \end{equation} 
 where $*$ represents the complex conjugate operation and ``vecdiag" is an operation which extracts and stacks the diagonal elements of a matrix into a vector. Also,  $\mathbf{\tilde{C}}(\mathbf{\boldsymbol{\xi}})|_{(t,t')}\in \mathbb{C}^{M \times M}$ represents the $(t,t')-$th sub-matrix of $\mathbf{\hat{C}(\boldsymbol{\xi})}$.

 \vspace{-2pt}
\section{CRLB Analysis}
\vspace{-1pt}
We derive the Fisher Information Matrix (FIM), followed by the analytical squared estimation error bound for evaluating the accuracy of the estimators in the previous section. 
\vspace{-2pt}
\subsection{FIM}
\label{FIM}
\vspace{-2pt}
 The derivation of FIM starts from equation \eqref{totaly}, according to \cite{b8}, the FIM  of the unknown vector $\boldsymbol{\xi}$ is given by
\begin{equation}
\begin{aligned} 
    \label{FIM}
    \mathbf{I}(\boldsymbol{\xi})_{i,j} &=\textrm{Tr}\Big[ \frac{\partial\hspace{1pt} \mathbf{C}(\boldsymbol{\xi})}{\partial\hspace{1pt}\boldsymbol{\xi}_i}\hspace{1pt}\mathbf{C}^{-1}(\boldsymbol{\xi})\hspace{1pt}\frac{\partial \hspace{1pt} \mathbf{C(\boldsymbol{\xi})}}{\partial \hspace{1pt}\boldsymbol{\xi}_j}\hspace{1pt}\mathbf{C}^{-1}(\boldsymbol{\xi})\Big] \\
  		&+ 2\Re\Big[\frac{\partial\hspace{1pt}\boldsymbol{\mu}^H(\boldsymbol{\xi})}{\partial\hspace{1pt}\boldsymbol{\xi}_i}\hspace{1pt}\mathbf{C}^{-1}(\boldsymbol{\xi})\hspace{1pt}\frac{\partial\hspace{1pt}\boldsymbol{\mu}(\boldsymbol{\xi})}{\partial\hspace{1pt}\boldsymbol{\xi}_j}\Big].\\
\end{aligned}
\end{equation}
\emph{We exercise a slight abuse of notation here when we denote the FIM as  $\mathbf{I}(\boldsymbol{\xi})$, since a $M\times{}M$ identity matrix is  denoted by $\mathbf{I}_M$}. We note that $\mathbf{I}(\boldsymbol{\xi})\in \mathbb{C}^{(2M+2)\times (2M+2)}$. Furthermore, $\textrm{Tr}[\cdot]$ denotes the matrix trace operator. According to \eqref{FIM}, the FIM $\mathbf{I}(\boldsymbol{\xi})$ is an addition of two matrices, namely, $\mathbf{I}(\boldsymbol{\xi})_{\mathbf{C}}$ and $\mathbf{I}(\boldsymbol{\xi})_{\boldsymbol{\mu}}$, where $[\mathbf{I}(\boldsymbol{\xi})_{\mathbf{C}}]_{i,j}$ 
is defined as the $(i,j)-$ th element of $\textrm{Tr}[(\partial\hspace{1pt} \mathbf{C}(\boldsymbol{\xi}))/(\partial\hspace{1pt}\boldsymbol{\xi}_i)\hspace{1pt}\mathbf{C}^{-1}(\boldsymbol{\xi})\hspace{1pt}(\partial\hspace{1pt} \mathbf{C}(\boldsymbol{\xi}))/(\partial\hspace{1pt}\boldsymbol{\xi}_j)\hspace{2pt}\hspace{1pt}\mathbf{C}^{-1}(\boldsymbol{\xi})]\hspace{1pt}$. Likewise,  
$\left[\mathbf{I}(\boldsymbol{\xi})_{\boldsymbol{\mu}}\right]_{i,j}$ is defined as the $(i,j)-$th element of $2\hspace{1pt}\Re[(\partial\boldsymbol{\mu}^H(\boldsymbol{\xi})/\partial\hspace{1pt}\boldsymbol{\xi}_i)\hspace{1pt}\mathbf{C}^{-1}(\boldsymbol{\xi})\hspace{1pt}(\partial\boldsymbol{\mu}(\boldsymbol{\xi})/\partial\hspace{1pt}\boldsymbol{\xi}_j)\hspace{1pt}]$. We first evaluate $\mathbf{I}(\boldsymbol{\xi})_\mathbf{C}$, which begins with calculating the \emph{derivative} of the $\mathbf{C}(\boldsymbol{\xi})$ with respect to elements in $\boldsymbol{\xi}$. That is, 
 \vspace{-2pt}
 \begin{equation}
 \hspace{0pt}
\label{Cmatrixderive}
    \frac{\partial\hspace{1pt} \mathbf{C(\boldsymbol{\xi})}}{\partial \hspace{1pt}\gamma} = 0 \hspace{8pt} \textrm{and} \hspace{8pt}\frac{\partial\hspace{1pt} \mathbf{C(\boldsymbol{\xi})}}{\partial\hspace{1pt} \sigma^2} =  \mathbf{\widetilde{I}}_T\otimes \mathbf{D}\mathbf{D}^{H}.  
    \vspace{-1pt}
\end{equation}
For every RF chain, $m = 1,2,\dots,\textit{M}$,
\vspace{1pt}
\begin{equation}
\label{Cd2}
\frac{\partial\hspace{1pt} \mathbf{C(\boldsymbol{\xi})}}{\partial\hspace{1pt} \alpha_m} = 0 \hspace{8pt} \textrm{and} \hspace{8pt}\frac{\partial\hspace{1pt} \mathbf{C(\boldsymbol{\xi})}}{\partial \hspace{1pt}d_m} = \widetilde{\mathbf{I}}_T\otimes\sigma^2 \mathbf{\widetilde{D}}_m.
\vspace{-3pt}
\end{equation}
Here $\mathbf{\widetilde{D}}_m = \textrm{diag}\{0,\dots,2d_m,\dots,0\}$ denotes a diagonal matrix.  Closely observing \eqref{Cmatrixderive} and \eqref{Cd2}, one can see that  $\mathbf{I}(\boldsymbol{\xi})_{\mathbf{C}}$ contains all \emph{zero} elements except for the sub-matrix blocks of $\mathbf{I(\boldsymbol{\xi})}_{\mathbf{C}}|_{(\sigma^2,\sigma^2)}$, $\mathbf{I(\boldsymbol{\xi})}_{\mathbf{C}}|_{(d,\hspace{1pt}d)}$, and $\mathbf{I(\boldsymbol{\xi})}_{\mathbf{C}}|_{(\sigma^2,\hspace{1pt}d)}$, respectively. To derive these three quantities, it is necessary to perform \emph{eigenvalue decompositions} of $\partial \hspace{1pt} \mathbf{C}(\boldsymbol{\xi})/\partial \hspace{1pt}\sigma^2$ and $\partial \hspace{1pt} \mathbf{C}(\boldsymbol{\xi})/\partial\hspace{1pt} d_m$ via $\tilde{\mathbf{Q}}$, which leads to the following representation:
\vspace{3pt}
\begin{equation}
\label{Csigma}
    \frac{\partial \hspace{1pt} \mathbf{C(\boldsymbol{\xi})}}{\partial \hspace{1pt} \sigma^2} = \tilde{\mathbf{Q}}^H\renewcommand*{\arraystretch}{.001}\begin{bmatrix}
    T\mathbf{D}\mathbf{D} ^H& & &\\
    & \hspace{-12pt} \boldsymbol{0} & &\\
    & & \hspace{-12pt}\ddots &\\
    & &  & \hspace{-12pt} \boldsymbol{0}
    \end{bmatrix}\tilde{\mathbf{Q}},
\end{equation}
and
\begin{equation}
\label{Cdm}
     \frac{\partial \hspace{1pt} \mathbf{C(\boldsymbol{\xi})}}{\partial \hspace{1pt}d_m} = \tilde{\mathbf{Q}}^H\renewcommand*{\arraystretch}{.01}\begin{bmatrix}
     \sigma^2 \mathbf{\widetilde{D}}_{m}T& & &\\
     & \hspace{-12pt} \boldsymbol{0} & &\\
     & & \hspace{-12pt}\ddots &\\
     & &  & \hspace{-12pt} \boldsymbol{0}
     \end{bmatrix}\tilde{\mathbf{Q}}.
     \vspace{-3pt}
\end{equation}
%Keep in mind that $\mathbf{C}\left(\boldsymbol{\xi}\right)=\tilde{\mathbf{Q}}^H\mathbf{\Lambda} \tilde{\mathbf{Q}}$. 
%According to \eqref{FIM}, \eqref{Cd2} and \eqref{Csigma}, 
Leveraging the \emph{unitary} property of $\tilde{\mathbf{Q}}$ and the \emph{cyclic} property of the $\textrm{Tr}[\cdot]$ operation, 
\begin{align}
    \nonumber
	 \left[\mathbf{I(\boldsymbol{\xi})}_{\mathbf{C}}\right]_{\sigma^2,\sigma^2} &= \textrm{Tr}\left[ \frac{\partial \hspace{1pt} \mathbf{C(\boldsymbol{\xi})}}{\partial \hspace{1pt}\sigma^2}\hspace{1pt}\mathbf{C}^{-1}(\boldsymbol{\xi}) \frac{\partial \hspace{1pt}\mathbf{C(\boldsymbol{\xi})}}{\partial\hspace{1pt}\sigma^2}\mathbf{C}^{-1}(\boldsymbol{\xi})\right] \\
	 & = \sum_{m=1}^{M}\frac{T^{\hspace{1pt}2}d_m^{\hspace{1pt}4}}{(\sigma^2Td_m^{\hspace{1pt}2}+N_0)^2}. \\[-17pt]
	 &\nonumber
\end{align}
Following a similar methodology, one can compute 
\vspace{1pt}
\begin{equation}
\label{calc1}
	 \left[\mathbf{I(\boldsymbol{\xi})}_{\mathbf{C}}\right]_{d_{m},d_{k}}= \frac{4\hspace{1pt}\sigma^4T^2d_m^{\hspace{1pt}2}}{(\sigma^2Td_k^2+N_0)^2}\delta_{mk},
	 \vspace{1pt}
\end{equation}	 
and	 
\begin{equation}	 
\label{calc2}
	 \left[\mathbf{I(\boldsymbol{\xi})}_{\mathbf{C}}\right]_{\sigma^2,d_m} = \frac{2\sigma^2T^2d_m^3}{(\sigma^2Td_m^2+N_0)^2},
	 \vspace{2pt}
\end{equation}
where $\delta_{mk}=1$ only if $m = k$. Due to space constraints, we avoid presenting 
the full calculation of \eqref{calc1} and \eqref{calc2}, respectively. Following this, we derive $\mathbf{I}(\boldsymbol{\xi})_{\boldsymbol{\mu}}$. We begin by taking the derivative of $\boldsymbol{\mu}(\boldsymbol{\xi})$ with respect to elements in $\boldsymbol{\xi}$. Doing this yields the following results 
   \begin{align}
   \nonumber
   \frac{\partial\hspace{1pt}\boldsymbol{\mu}(\boldsymbol{\xi})}{\partial\hspace{1pt}\gamma} &= \mathbf{1} \otimes \mathbf{D}\hspace{1pt}\mathbf{a}(\phi),
    \hspace{10pt}
    \frac{\partial\hspace{1pt}\boldsymbol{\mu}(\boldsymbol{\xi})}{\partial \hspace{1pt}d_m} = \mathbf{1} \otimes e^{j\alpha_m} \mathbf{E}_{mm}\gamma\hspace{2pt}\mathbf{a}(\phi),\\ \label{CRLBmu}
    \frac{\partial\hspace{1pt}\boldsymbol{\mu}(\boldsymbol{\xi})}{\partial\hspace{1pt}\sigma^2} &\hspace{-2pt}=\hspace{-2pt}0,
    \hspace{6pt} \textrm{and} \hspace{6pt}
    \frac{\partial\hspace{1pt}\boldsymbol{\mu}(\boldsymbol{\xi})}{\partial \hspace{1pt}\alpha_m}\hspace{-1pt}=\hspace{-1pt}\mathbf{1} \otimes jd_me^{j\alpha_m} \mathbf{E}_{mm}\gamma\hspace{1pt}\mathbf{a(\phi)}.
   \end{align}
Note that $\mathbf{E}_{mm}$ is the elementary matrix which has unit value only at the intersection of the $m-$th row and $m-$th column, and zeros elsewhere. In accordance with \eqref{CRLBmu}, it is trivial that $\mathbf{I(\boldsymbol{\xi})}_{\boldsymbol{\mu}}|_{(\sigma^2, d)}$, $\mathbf{I(\boldsymbol{\xi})}_{\boldsymbol{\mu}}|_{(\sigma^2, \alpha)}$,
$[\mathbf{I(\boldsymbol{\xi})}_{\boldsymbol{\mu}}]_{\sigma^2, \hspace{1pt}\sigma^2}$,
and $[\mathbf{I(\boldsymbol{\xi})}_{\boldsymbol{\mu}}
]_{\sigma^2, \gamma}$ are all 0, since the first two quantities are zero vectors, while the second two quantities are zero scalars. To derive the remaining sub-matrices of $\mathbf{I}(\boldsymbol{\xi})_{\boldsymbol{\mu}}$, we express a unit vector as $\mathbf{1} = \sqrt{T}\mathbf{Q}^H\mathbf{\boldsymbol{\eta}}$, where $\mathbf{\boldsymbol{\eta}}$ denoted as a $T\times{}1$ column vector $[1,0,\dots,0]^T$. Based on the properties of the unitary matrix and the mixed-product property of the kronecker operation,  we can express  $[\mathbf{I(\boldsymbol{\xi})}_{\boldsymbol{\mu}}]_{d_m, d_k}$ as
   \begin{align}
   \nonumber
  	[\mathbf{I(\boldsymbol{\xi})}_{\boldsymbol{\mu}}]_{d_m, d_k} &= 2\hspace{1pt}\Re\left[\frac{\partial\hspace{1pt}\mathbf{\boldsymbol{\mu}}^H\boldsymbol{(\xi)}}{\partial\hspace{1pt} d_m}\hspace{2pt}\mathbf{C^{-1}(\boldsymbol{\xi})}\hspace{2pt}\frac{\partial\hspace{1pt}\boldsymbol{\mu(\xi)}}{\partial \hspace{1pt}d_k}\right] \\[5pt] \nonumber
  	& \overset{(a)}= 2\hspace{1pt}\Re\left[\boldsymbol{\kappa}^H(\mathbf{Q} \otimes \mathbf{I}_M)^{-1}
  \mathbf{\Lambda}^{-1}(\mathbf{Q} \otimes \mathbf{I}_M)\boldsymbol{\kappa}\right] \\[5pt]
  	& = \frac{2\hspace{1pt}T}{N_0+\sigma^2\hspace{1pt}T\hspace{1pt}d^2m}\hspace{1pt}\gamma^2
  	\hspace{1pt}\delta_{mk}, \\[-15pt]
  	&\nonumber
   \end{align}
where $(a)$ contains $\boldsymbol{\kappa}=\mathbf{Q}^H\mathbf{\boldsymbol{\eta}}\otimes \gamma \hspace{1pt}e^{j\hspace{1pt}\alpha_m} \mathbf{E}_{mm}\hspace{1pt}\mathbf{a(\phi)}$. Following the same method, we can derive the rest of sub-matrices $\mathbf{I(\boldsymbol{\xi})}_{\boldsymbol{\mu}}$. Due to space limitation, we avoid presenting the exact calculations, however we quote the final results below:
\vspace{-3pt}
\begin{equation}
\hspace{-31pt}
[\mathbf{I(\boldsymbol{\xi})}_{\boldsymbol{\mu}}]_{\gamma,\hspace{1pt}\gamma}= \sum_{m=1}^{M}\frac{2\hspace{1pt}d_m^{\hspace{1pt}2}\hspace{1pt}T}{\sigma^2\hspace{1pt}
T\hspace{1pt}d^{\hspace{1pt}2}_m+N_0},
\end{equation}
\begin{equation}
\hspace{-47pt}
[\mathbf{I(\boldsymbol{\xi})}_{\boldsymbol{\mu}}]_{d_m,\hspace{1pt}\gamma}= \frac{2\hspace{1pt}T\hspace{1pt}d_m\gamma}{\sigma^2Td_m^2+N_0},
\vspace{2pt}
\end{equation}
\begin{equation}
\hspace{-23pt}
\label{23}
[\mathbf{I(\boldsymbol{\xi})}_{\boldsymbol{\mu}}]_{\alpha_m,\hspace{1pt}\alpha_k}
=\frac{2\hspace{1pt}T\hspace{1pt}d^2_m\gamma^2}{N_0+\sigma^2\hspace{1pt}T\hspace{1pt}d^2_m}
\delta_{m\hspace{1pt}k}, 
\vspace{2pt}
\end{equation}
\begin{equation}
[\mathbf{I(\boldsymbol{\xi})}_{\boldsymbol{\mu}}]_{\gamma,\hspace{1pt}\alpha_k}=0\hspace{8pt}
\textrm{and}\hspace{8pt}
[\mathbf{I(\boldsymbol{\xi})}_{\boldsymbol{\mu}}]_{d_m,\hspace{1pt}\alpha_k}=0.
\vspace{1pt}
\end{equation}
Adding $\mathbf{I(\boldsymbol{\xi})}_{\boldsymbol{\mu}}$ with $\mathbf{I(\boldsymbol{\xi})}_{\mathbf{C}}$, the \emph{closed-form} FIM $\mathbf{I(\boldsymbol{\xi})}$ is given by \eqref{FIMmatrix}, presented on top of the following page. 
\begin{figure*}[!t]
\begin{equation}
\label{FIMmatrix}
	\mathbf{I}(\boldsymbol{\xi}) = \begin{bmatrix}
	\mathbf{I(\boldsymbol{\xi})}_{\boldsymbol{\mu}}|_{(d,d)}+\mathbf{I(\boldsymbol{\xi})}_{\mathbf{C}}|_{(d,d)} 	\hspace{-6pt}& \mathbf{I(\boldsymbol{\xi})}_{\mathbf{C}}|_{(\sigma^2,d)} & \hspace{-8pt}\boldsymbol{0} &\hspace{-10pt} \mathbf{I(\boldsymbol{\xi})}_{\boldsymbol{\mu}}|_{(d,\gamma)} \\
	 \mathbf{I(\boldsymbol{\xi})}_{\mathbf{C}}^H|_{(\sigma^2,d)} 	\hspace{-8pt}&  \mathbf{I(\boldsymbol{\xi})}_{\mathbf{C}}|_{(\sigma^2,\sigma^2)} & \hspace{-8pt}\boldsymbol{0} & \hspace{-8pt} \boldsymbol{0}\\
	 \boldsymbol{0} 	\hspace{-8pt}& \boldsymbol{0} & 	\hspace{-8pt}\mathbf{I(\boldsymbol{\xi})}_{\boldsymbol{\mu}}|_{(\alpha,\alpha)} &	\hspace{-8pt}\mathbf{I(\boldsymbol{\xi})}_{\boldsymbol{\mu}}|_{(\alpha,\gamma)}\\
	 \mathbf{I(\boldsymbol{\xi})}_{\boldsymbol{\mu}}^H|_{(d,\gamma)}	\hspace{-8pt} &\boldsymbol{0} &  \hspace{-8pt}\mathbf{I(\boldsymbol{\xi})}_{\boldsymbol{\mu}}^H|_{(\alpha,\gamma)} & \hspace{-8pt}\mathbf{I(\boldsymbol{\xi})}_{\boldsymbol{\mu}}|_{(\gamma,\gamma)}
	\end{bmatrix}.
\end{equation}
\vspace{4pt}
\hrulefill
\vspace{-14pt}
\end{figure*}
\vspace{-3pt}
\subsection{Inverse of FIM}
\vspace{-2pt} 
To compute the CRLB, one is generally required to \emph{invert} the FIM. We check the invertability of $\mathbf{I}(\boldsymbol{\xi})$ by computing its determinant numerically, and ensuring that the result is non-zero. In order to perform parameter estimation for absolute calibration, we are only interested in the following two terms of the FIM: $\mathbf{I(\boldsymbol{\xi})}_{d,d}^{-1}$ and $\mathbf{I(\boldsymbol{\xi})}_{\alpha,\alpha}^{-1}$. \emph{This is since only these terms contain the necessary information for the amplitude scaling and phase shifts associated with each RF chain. The other terms do not need to be inverted, since they contain information relating to $\gamma$ and $\sigma^2$ which denote the LOS power and power of the diffuse multipath components which do not need to be estimated.} Enlightened by this, we provide the following analysis which begins by splitting $\mathbf{I(\boldsymbol{\xi})}$ into four parts for mathematical convenience. Specifically, 
\begin{equation}
	\mathbf{I(\boldsymbol{\xi})} = \begin{bmatrix}
	   \mathbf{X} & \boldsymbol{\psi}\\
	   \boldsymbol{\psi}^{H} & w\\
	\end{bmatrix},
\end{equation}
where the scalar $w = \mathbf{I(\boldsymbol{\xi})}_{\boldsymbol{\mu}}|_{(\gamma,\gamma)}$, the vector $\boldsymbol{\psi}$ is given by  
$(\mathbf{I(\boldsymbol{\xi})}^T_{\boldsymbol{\mu}}|_{(d,\gamma)}\hspace{7pt} \boldsymbol{0}_{1\times M} \hspace{7pt} \mathbf{I(\boldsymbol{\xi})}^T_{\boldsymbol{\mu}}|_{(\alpha,\gamma)})^T$ and $\mathbf{X}$ for the rest of $\mathbf{I(\boldsymbol{\xi})}$. 
Leveraging the relationship between the \emph{adjugate matrix} and the \emph{inversion matrix},  $\mathbf{I}^{-1}(\boldsymbol{\xi})$ can be expressed as 
\begin{equation}
	\mathbf{I}^{-1}(\boldsymbol{\xi}) = \frac{\mathbf{I}^\dagger(\boldsymbol{\xi})}{\det(\mathbf{I}(\boldsymbol{\xi}))},
\end{equation}
where $\mathbf{I^\dagger(\boldsymbol{\xi})}$ is the adjugate matrix, which can be obtained by extracting the resulting sub-matrix after striking out the $i-$th row and column of $\mathbf{I(\boldsymbol{\xi})}$. Since only the diagonal elements of $\mathbf{X}$ contain the phase shift and amplitude scaling estimation parameters of interest, the range of $i=1,2,\dots,M,M+2,M+3,...,2M+1$. Then, by applying the definition of adjugate matrix and \emph{Schur complement}, one can calculate the $i-$th diagonal element of $\mathbf{I^{-1}(\boldsymbol{\xi})}$ as \cite{matrix}
\vspace{-1pt}
\begin{equation}
\label{28}
 \mathbf{I}^{-1}(\boldsymbol{\xi})_{ii}= \frac{\det(\mathbf{\widetilde{X}}_{ii})}{\det(\mathbf{X})}\left(\frac{w-\boldsymbol{\eta}_i^H\mathbf{\widetilde{X}}^{-1}_{ii}\boldsymbol{\eta}_i}{w-\boldsymbol{\eta}^H\mathbf{X}^{-1}\boldsymbol{\eta}}
 \right), 
 \vspace{-6pt}
\end{equation}
where $\mathbf{\widetilde{X}}_{ii}$ can be obtained $\mathbf{X}$ by striking the $i-$th row and column, while $\boldsymbol{\eta}_{i}$ can be extracted from the column vector $\boldsymbol{\eta}$ by striking the $i$-th row. We now present the full analytical form of $\mathbf{I}^{-1}(\boldsymbol{\xi})_{ii}$: For convenience, we let vectors $\boldsymbol{\phi}\in \mathbb{R}^M$, $\boldsymbol{\zeta} \in \mathbb{R}^M$ denote the diagonal elements of the matrices  $\mathbf{I}(\boldsymbol{\xi})_{\boldsymbol{\mu}}|_{(d,d)}+\mathbf{I}(\boldsymbol{\xi})_{\mathbf{C}}|_{(d,d)}$ and $\mathbf{I}(\boldsymbol{\xi})_{\boldsymbol{\mu}}|_{(\alpha,\alpha)}$ respectively. Furthermore, we let $\boldsymbol{\varphi} \in \mathbb{R}^M$ and  $\boldsymbol{\vartheta}\in \mathbb{R}^{M}$ represent vectors $\mathbf{I(\boldsymbol{\xi})}_{\mathbf{C}}|_{(\sigma^2,d)}$ and $\mathbf{I(\boldsymbol{\xi})}_{\boldsymbol{\mu}}|_{(d,\gamma)}$ respectively. In addition, we define scalars $\rho=\mathbf{I(\boldsymbol{\xi})}_{\mathbf{C}}|_{(\sigma^2,\sigma^2)}$, $\chi_i=\det(\mathbf{\widetilde{X}}_{ii})/\det(\mathbf{X})$ and $\chi_i' = (w-\boldsymbol{\eta}_i^H\mathbf{\widetilde{X}}^{-1}_{ii}\boldsymbol{\eta}_i)/(w-\boldsymbol{\eta}^H\mathbf{X}^{-1}\boldsymbol{\eta})$. With the aid of \emph{Gaussian elimination}, we can calculate $\chi_i$ as \cite{matrix}
\begin{equation}
\label{inverse1}
	\chi_i= \left\{
	\begin{aligned}
	 \frac{\rho-\sum_{j \neq i}^{M}\frac{\boldsymbol{\varphi}_j^2}{\boldsymbol{\phi}_j}}{\boldsymbol{\phi}_i(\rho-\sum_{j=1}^ M\frac{\boldsymbol{\varphi}_j^2}{\boldsymbol{\phi}_j})} \hspace{1pt};\hspace{10pt} i = 1,2,...,M\\
	 \frac{1}{\boldsymbol{\zeta}_i} \hspace{1pt}; \hspace{10pt}i = M+2,M+3,...,2M+1. 
	\end{aligned}
	\right.
\end{equation}
%and $\boldsymbol{\zeta} \in \mathbb{R}^M$
 Using block matrix inversion theorem \cite{matrix}, 
\begin{equation}
\label{inverse2}
\hspace{-3pt}\chi_i'= \left\{
\begin{aligned}
\frac{w-\big(\sum_{j\neq i}^{M}\frac{\boldsymbol{\vartheta}_j^2}{\boldsymbol{\phi}_j}\big)-f_2^{-1}\big(\sum_{j\neq i}^{M}\frac{\boldsymbol{\vartheta}_j\hspace{1pt}\boldsymbol{\varphi}_j}{\boldsymbol{\phi}_j}\big)^2}
{w-\big(\sum_{j=1}^{M}\frac{\boldsymbol{\vartheta}_j^2}{\boldsymbol{\phi}_j}\big)-f_1^{-1}\big(\sum_{j=1}^{M}\frac{\boldsymbol{\vartheta}_j\hspace{1pt}\boldsymbol{\varphi}_j}{\boldsymbol{\phi}_j}\big)^2}
\\ \hspace{22pt};i = 1,2,...,M\\ 
1\hspace{15pt}; \hspace{2pt} i = M+2,M+3,...,2M+1,
\end{aligned}
\right.
\end{equation}
where scalrs $f_1$ and $f_2$ are defined as $f_1 = \rho-\sum_{j=1}^ M(\boldsymbol{\varphi}_j^2/\boldsymbol{\phi}_j)$, $f_2 = \rho-\sum_{j \neq i}^{M}(\boldsymbol{\varphi}_j^2/\boldsymbol{\phi}_j)$, respectively. 

Note that this is a very 
general solution to a complex problem which holds for any SNR value and any number of receive antennas at the BS. An interesting special case of \eqref{inverse2} can be analyzed, which is as follows: Supposing that the system is operated in the high SNR regime, implying that $N_0$ is much \emph{smaller} than $\sigma^2 \hspace{1pt} T \hspace{1pt} d^2_i$, when $1\leq i\leq M$, $\chi_i'$ can be approximated as:
\begin{equation}
    \label{chispecialcase}
	\chi_i'\hspace{2pt}\approx\hspace{2pt}\frac{\varepsilon\hspace{1pt}\big[ \sigma^2M\hspace{1pt}(2\hspace{1pt}\varepsilon-\gamma^2)+2\hspace{1pt}\sigma^4+\gamma^2\hspace{1pt}\varepsilon+2\hspace{1pt}\gamma^2\hspace{1pt}\sigma^2\big]}{\sigma^2\hspace{1pt}(2\hspace{1pt}\varepsilon-\gamma^2)\hspace{1pt}(M\hspace{1pt}\varepsilon+\sigma^2)},
\end{equation}
where $\varepsilon$ is defined as $\varepsilon = 2\hspace{1pt}T\hspace{1pt}\gamma^2+(4\hspace{1pt}T^2-1)\hspace{1pt}\sigma^2$. If $M$ is much larger than $\sigma^2$, which is typically the case for massive MIMO systems, then $\chi_i'$ can be approximated as 1 since the numerator and denominator of \eqref{chispecialcase} both scale linearly with $M$ resulting in a cancellation. Relative to $M$, the other variables do not significantly influence the result of \eqref{chispecialcase} and hence are less dominant. Based on \eqref{28}-\eqref{inverse2}, in high SNR conditions, for a massive MIMO system, the diagonal elements of $\mathbf{I^{-1}(\boldsymbol{\xi})}$ can be revealed in a rather elegant form.  which demonstrate the CRLBs of the amplitude and phase estimations. These are 
\begin{equation}
\label{finalinv}
	\mathbf{I}^{-1}(\boldsymbol{\xi})_{ii} \approx \left\{
	\begin{aligned}
	\frac{\sigma^2d_i^2}{2\hspace{1pt}(\gamma^2+2\sigma^2)} \hspace{37pt}; i = 1,2,...,M\\ 
	\frac{\sigma^2}{2\hspace{1pt}\gamma^2} \hspace{5pt}; i = M+2,M+3,...,2M+1.
	\end{aligned}
	\right.
\end{equation}
\emph{From \eqref{finalinv}, the CRLB of the phase estimation is proportional to $\sigma^2$ and inversely proportional to $\gamma^2$. As a special case, if the system is operating with pure non LOS propagation environment (i.e., $\gamma = 0$), the CRLB relating to phase drifts goes to infinity, and it is impossible to estimate the phase drifts in this situation. However, for UE-aided absolute calibration, it is common that the UE will be in close proximity to the BS and hence will almost surely have a dominant LOS component, along with other multipath components. In contrast, for the amplitude estimations, both $\gamma$ and $\sigma^2$ can contribute to the inverse of the FIM. Therefore, it is possible to find a soluable estimator, even when there is no LOS component.}

%Numerical Results
\vspace{-2pt}
\section{Numerical Results}
\vspace{-2pt}
The ultimate aim of our work is to implement the aforementioned calibration parameter estimation techniques into a real-time massive MIMO testbed. To this end, as a first step in this direction, we evaluate the estimation performance via Monte-Carlo simulations. Our simulation framework caters to a  100 element ULA connected to 100 individual RF chains. We assume that the physical distance between the electrical phase centers of successive antenna elements is $d=\lambda_{f}/2$, where $\lambda_{f}$ is the wavelength corresponding to the operating carrier frequency. Considering this, the overall steering vector can be written as
\begin{equation}
    \label{steering vector}
    \mathbf{a}\left(\phi\right)=\left[1,e^{-j2\pi{}d\cos(\phi)}
    \dots,e^{-j2\pi{}d(M-1)\cos(\phi)}\right]. 
\end{equation}
Consistent with \cite{R5}, the \emph{ground truth} of the \emph{magnitudes} of the RF chain coefficients are assumed to be unity, while the \emph{phases} are assumed to be distributed uniformly between $[-\pi,\pi]$. This serves as a basis for comparison for the estimated amplitudes and phases. Furthermore, the SNR is given by 
    \begin{align}
        \nonumber
        \textrm{SNR} &= \frac{\mathbb{E}\hspace{1pt}\{(\gamma\mathbf{D}\mathbf{a}(\phi)\hspace{1pt}p+p
        \hspace{1pt}\mathbf{D}\mathbf{h})(\gamma\mathbf{D}\mathbf{a}(\phi)p + \mathbf{D}\mathbf{h}p)^H\}}
        {E\left\{\mathbf{n}\mathbf{n}^{H}\right\}}  \\ 
        \label{SNR}
        &= \frac{\sum_{m=1}^{M}d_m^2(\sigma^2+\gamma^2)}{MN_0}, \\[-20pt]
        &\nonumber
    \end{align}  
\begin{figure}[!t]
  \hspace{-10pt}
  	\includegraphics[width=8.7cm]{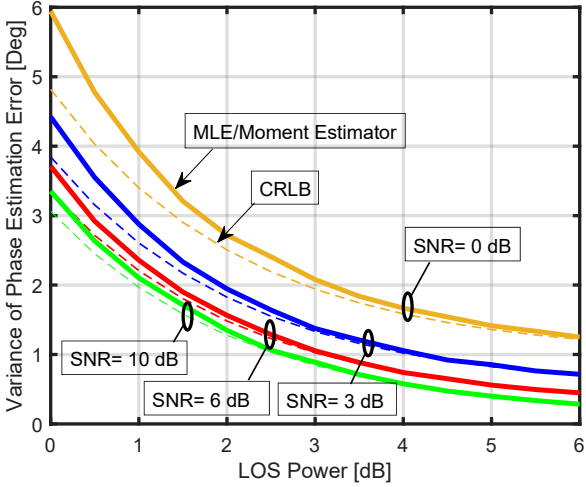}
  	\vspace{-5pt}
  	\caption{MLE (and moment-based) estimator performance as a function of LOS powers for phase calibration with varying SNRs. The phase estimation CRLBs are shown for comparison purposes.}
  	\vspace{-18pt}
  	\label{MLE}
  \end{figure}
where the respective quantities in \eqref{SNR} are defined in Sec.~\ref{SystemModelS}. To manage the computational run-time of the numerical simulations, while observing data for a long enough time period, we set the total observation duration of the received vector as $T = 3$, for each 10 independent and identically distributed Monte-Carlo realizations are simulated. To evaluate the accuracy of phase and amplitude estimations, we hereby assume that the groundtruths of the phase drifts and amplitude scalings of each RF chain are generated by the simulation framework, and stored for the sake of comparison. In Fig.~\ref{MLE}, we first present the performance of \emph{both} types of \emph{phase} estimators (mathematically proven to have same form), with the derived CRLB for phase estimation. We do this by reporting the variance of the phase MLE estimators over the 100 parallel RF chains against the derived CRLB under different SNRs and LOS powers ($\gamma$) factors. The CRLB of the phase estimation is inversely proportional to $\gamma^2$ (see \eqref{23}), thus the CRLB and the MLE estimator decay exponentially as $\gamma$ increases. As shown, the variances of the MLE estimator approaches the CRLB as $\gamma$ or the SNR increases. The results coincide with the MLE behaviour since the increase of either the SNR or LOS power reduces the phase variance, leading to a small estimation error and the MLE is therefore able to achieve its asymptotic probability distribution function \cite{b8}. Naturally, we would expect the moment-based estimator to have the same performance via \eqref{momentbasedestresult} and \eqref{Esteq}. 

\begin{figure}[!t]
  \vspace{-3pt}
  \hspace{-5pt}
  	\includegraphics[width=8.9cm]{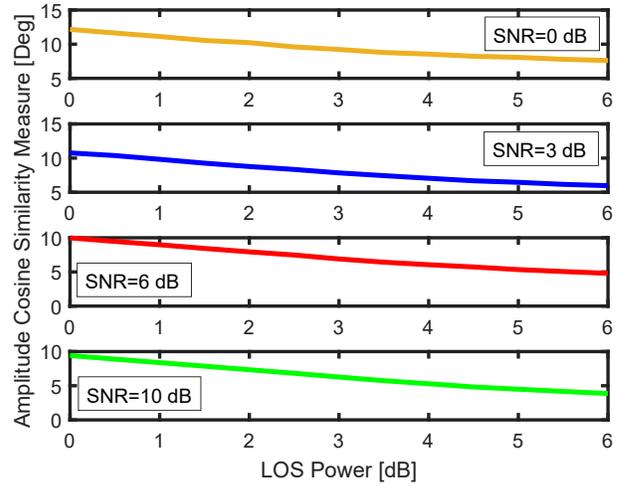}
  	\vspace{-20pt}
  	\caption{Cosine similarity measure as a function of the LOS powers for amplitude estimation using the moment-based estimator. Variability in SNR is also presented.}
  	\label{MM}  
  	\vspace{-18pt}
  \end{figure}

Figure~\ref{MM} presents the estimation result of the linear space spanned by the amplitude vector $\mathbf{d} = [d_1, d_2,...,d_M]^T$ using the moment-based estimator. Using the \emph{cosine similarity measure} \cite{b10}, we define the criteria for measuring the angular difference between the estimated vector, $\hat{\mathbf{d}}$, and the groundtruth, $\mathbf{d}$, as
\vspace{-5pt}
\begin{equation}
\label{cossimilarity}
    \textrm{cos. sim.} = \mathrm{arccos}\left(\frac{|\hspace{1pt}\hat{\mathbf{d}}^H\mathbf{d}\hspace{1pt}|}
    {\|\hat{\mathbf{d}}\|\hspace{2pt}\|\mathbf{d}\|}\right), 
\end{equation}
where $\|\cdot\|$ denotes the vector norm. Based on \eqref{cossimilarity}, a perfectly estimated vector is aligned with the groundtruth vector, i.e., $\mathrm{cos. sim.}=0$, while the worst estimate is a vector in perpendicular direction to the groundtruth vector resulting in $\mathrm{cos. sim.}=1$. The moment-based estimator in \eqref{mom} exploits the first and 
second-order statistics by calculating the expectation of the received signal and reconstructing the covariance matrix. The increase of LOS power yields more superior reconstruction quality and therefore improves the estimation result. Also, as shown in Fig.~\ref{MM}, an increase in the SNR results in higher estimator accuracy as the estimated amplitude starts to converge towards the groundtruth. In addition, it is challenging to evaluate the CRLB of  \eqref{cossimilarity}, since the inversion of the whole FIM is required, which is an extremely difficult task \cite{b8}. Therefore, we defer this to the upcoming journal version of the paper. 
%\begin{figure}[h]
%  \centering
%      \hspace{-10pt}\includegraphics[width=0.55\textwidth]{Lumami2modes_phase.eps}
%  \caption{The phase estimation results of two Lumami Running modes}
%\end{figure}
%\begin{figure}[h]
%  \centering
%      \hspace{-10pt}\includegraphics[width=0.55\textwidth]{Lumami2modes_amplitude.eps}
%  \caption{The amplitude estimation results of two Lumami Running modes}
%\end{figure}
%Figure 5 and figure 6 present the amplitude and phase estimation results regarding to two operating modes of Lumami. As illustrated, the proposed phase and amplitude estimation algorithms lead to almost the same calibration quality since each elements of the steering vectors has constant amplitude as one, therefore the power of the line of sight part remain the same regarding to both of these two operation modes. Therefore, the estimation results are the same to each other.   
\begin{figure}[!t]     
\vspace{-10pt}
  \centering
    \hspace{-10pt}
     \includegraphics[width=8.93cm]{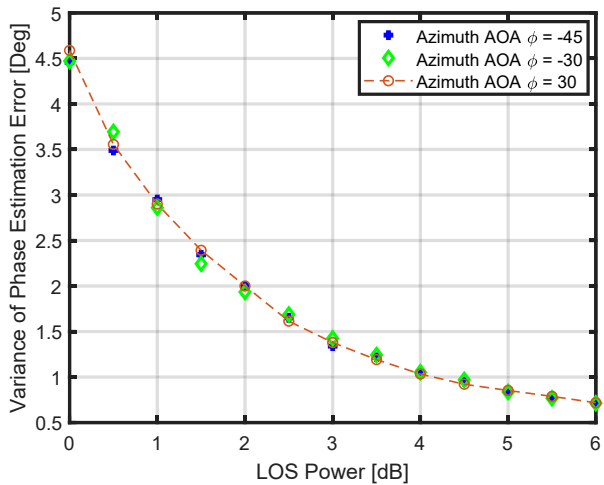}
     \vspace{-10pt}
 \caption{MLE (and moment-based) estimator performance for different LOS 
 powers and dominant LOS AOAs (in degrees) with SNR=3 dB.}
 \vspace{-18pt}
 \label{DiffAngles}
\end{figure}

Figure~\ref{DiffAngles} depicts the phase estimation results of the two estimators as a function of LOS powers, for different dominant LOS AOAs at SNR=3 dB. It can readily be observed that the resulting phase estimates for different AOAs are essentially the same. This is since each element of the steering vector has a constant amplitude of one for all of the incoming angles between $[-\pi/2, \pi/2]$. To this end, the change of dominant AOAs will have no influence on the estimator performance for a given LOS power and a given SNR. Although not shown here, the same trends hold for the moment-based amplitude estimator following the same phenomena.

%Conclusion
\vspace{-4pt}
\section{Conclusion}
\label{Conclusion}
\vspace{-1pt}
In this paper, we consider UE-aided amplitude and phase estimation for absolute calibration of massive MIMO front-ends. Assuming a Ricean fading channel model, for a single-user massive MIMO system, we analyze the performance of moment-based and MLE estimators for estimating the relative amplitude scalings and phase shifts associated with each RF chain. Our analysis assumes no knowledge of the LOS power, diffuse multipath component power, amplitudes and phase shifts. The derived estimators only need knowledge of the received array's steering vectors and transmitted pilots by the UE. We mathematically prove that for phase estimation, both MLE and moment-based estimators have the same form, and hence perform equally well. To evaluate the performance of respective estimators, we investigate the CRLB via the analysis of the FIM, where we draw several important insights. We show that the presence of a dominant LOS component is mandatory for phase drift estimation, while not necessary for the amplitude estimation. Our numerical results indicate that the variance of the phase estimates converge to the corresponding CRLBs with increasing LOS powers and SNRs. Likewise, the amplitude estimation accuracy improves substantially with increasing SNR. Different dominant LOS AOA tends to make almost no difference to the phase and amplitude estimator performance, since the amplitude of each entry in the steering vector remains constant over all considered angles. In the future, our aim is to implement the estimators for absolute calibration on the Lund University massive MIMO testbed.

%Acknowledgment
\vspace{-5pt}
\section*{Acknowledgment}
\vspace{-3pt}
The authors gratefully acknowledge Prof. Bo Bernhardsson, A/Prof. Fredrik Rusek and Mr. Tianyi Bai. The support of Ericsson AB is also acknowledged for funding this work.

%References
      
\vspace{12pt}

\vspace{-6pt}
\begin{thebibliography}{00}
\vspace{-4pt}
\bibitem{ERICSSON1}Ericsson, ``Ericsson Mobility Report," June 2019, Accessible online at https://www.ericsson.com/49d1d9/assets/local/mobility-report. 
\bibitem{3GPPTR38104}3GPP TR 38.104, ``Technical Specification on base station (BS) radio transmission and reception," v16.1.0, Sep. 2019.  
\bibitem{MARZETTA1}T.~L.~Marzetta, ``Noncooperative cellular wireless with unlimited numbers of base station antennas," \emph{IEEE Trans. Wireless Commun.}, vol. 9, no. 11, pp. 3590-3600, Nov. 2010.
\bibitem{LARSSON1} E.~G.~Larsson, O.~Edfors, F.~Tufvesson, and T.~L.~Marzetta, ``Massive MIMO for next generation wireless systems," \emph{IEEE Commun.Mag.}, vol. 52, no. 2, pp. 186-195, Feb. 2014.
\bibitem{YANG1} H.~Yang and T.~L.~Marzetta, ``Performance of conjugate and zero-forcing beamforming in large-scale antenna systems," \emph{IEEE J. Sel. Areas Commun.}, vol. 31, no. 2, pp. 172-179, Feb. 2013.
\bibitem{GAO1} X.~Gao, O.~Edfors, F.~Tufvesson, and E.~G.~Larsson, ``Massive MIMO in real propagation environments: Do all antennas contribute equally?," \emph{IEEE Trans. Commun.}, vol. 63, no. 11, pp. 3917-3928, Nov. 2015.
\bibitem{SHAFI1}M.~Shafi, \emph{et al.}, ``5G: A tutorial overview of standards, trials, challenges, deployment, and practice,'' \emph{IEEE J. Sel. Areas Commun.}, vol. 35, no. 6, pp. 1201-1221, Jun. 2017. 
\bibitem{SHAFI2}M.~Shafi, \emph{et al.}, ``Microwave vs. millimeter-wave propagation channels: Key differences and impact on 5G cellular systems," \emph{IEEE Commun. Mag.}, vol. 56, no. 12, pp. 14-20, Dec. 2018. 
\bibitem{MOLISCH1}A.~F.~Molisch, \emph{et al.}, ``Hybrid beamforming for massive MIMO: A survey," \emph{IEEE Commun. Mag.}, vol. 55, no. 9, pp. 134-141, Sep. 2017. 
\bibitem{R1} M. ~Guillaud, D.~T.~M.~Slock and R.~Knopp, ``A practical method for wireless channel reciprocity exploitation through relative calibration," \emph{Proc. 8th Int. Symp. Signal Processing and Its Applications (ISSPA)}, 2005., Sydney, Australia, 2005, pp. 403-406.
\bibitem{R2} R.~Rogalin, \emph{et al.}, ``Scalable synchronization and reciprocity calibration for distributed multiuser MIMO," \emph{IEEE Trans. Wireless Commun.}, vol. 13, no. 4, pp. 1815-1831, Apr. 2014.
\bibitem{R3} J.~Vieira, F.~Rusek, and F.~Tufvesson, ``Reciprocity calibration methods for massive MIMO based on antenna coupling," \emph{Proc. IEEE Global Commun. Conf. (GLOBECOM)}, 2014, pp. 3708-3712.
\bibitem{R4} J.~Vieira, F.~Rusek, O.~Edfors, S.~Malkowsky, L.~Liu, and F.~Tufvesson, ``Reciprocity calibration for massive MIMO: Proposal, modeling, and validation," \emph{IEEE Trans. Wireless Commun.}, vol. 16, no. 5, pp. 3042-3056, May 2017.
\bibitem{R5} X.~Jiang \emph{et al.}, ``A framework for over-the-air reciprocity calibration for TDD massive MIMO systems," \emph{IEEE Trans. Wireless Commun.}, vol. 17, no. 9, pp. 5975-5990, Sept. 2018.
\bibitem{A1} B.~C.~Ng and C.~M.~S.~See, ``Sensor-array calibration using a maximum-likelihood approach," \emph{IEEE Trans. Antennas Propag.}, vol. 44, no. 6, pp. 827-835, Jun. 1996.
\bibitem{A2} H.~Liu, L.~Zhao, Y.~Li, X.~Jing, and T.~Truong, "A sparse-based approach for DOA estimation and array calibration in uniform linear array," \emph{IEEE Sensors J.}, vol. 16, no. 15, pp. 6018-6027, Aug. 2016.
\bibitem{A3} X.~Luo, F.~Yang, and H.~Zhu, "Massive MIMO self-calibration: Optimal interconnection for full calibration," \emph{IEEE Trans. Veh. Technol.} (early access), Sep. 2019.  
\bibitem{MOLISCH1}A.~F.~Molisch, \emph{Wireless Communications}, IEEE Wiley Press, 2011. 
\bibitem{TSE1}D.~Tse and P.~Viswanath, \emph{Fundamentals of Wireless Communication}, Cambridge University Press, 2005. 
\bibitem{GOLDSMITH1}A.~Goldsmith, \emph{Wireless Communications}, Cambridge University Press, 2005. 
\bibitem{A4} A.~Benzin and G.~Caire, ``Internal self-calibration methods for large scale array transceiver software-defined radios," \emph{21th International ITG Workshop on Smart Antennas (WSA)}, Mar. 2017, pp. 49-56.
\bibitem{A5} I.~Şeker, ``Calibration methods for phased array radars," in \emph{SPIE 8714, Radar Sensor Technology XVII, 87140W}, May. 2013. 
\bibitem{A6} Z.~Liu, ``Conditional Cramer–Rao lower bounds for DOA estimation and array calibration," \emph{IEEE Signal Process. Lett.}, vol. 21, no. 3, pp. 361-364, Mar. 2014. 
\bibitem{Intra} H.~Wei, D.~Wang, H.~Zhu, J.~Wang, S.~Sun, and X.~You, ``Mutual coupling calibration for multiuser massive MIMO systems," \emph{IEEE Trans. Wireless Commun.}, vol. 15, no. 1, pp. 606-619, Jan. 2016.
\bibitem{TATARIA1}H.~Tataria, P.~J.~Smith, L.~J.~Greenstein, P.~A.~Dmochowski, and M.~Matthaiou, ``Impact of line-of-sight and unequal spatial correlation on uplink MU-MIMO systems," \emph{IEEE Wireless Commun. Lett.}, vol. 6, no. 5, pp. 634-637, Oct. 2017. 
\bibitem{TATARIA2}H.~Tataria, P.~J.~Smith, L.~J.~Greenstein, and P.~A.~Dmochowski, ``Zero-forcing precoding performance in multiuser MIMO systems with heterogeneous Ricean fading," \emph{IEEE Wireless Commun. Lett.}, vol. 6, no. 1, pp. 74-77, Nov. 2016. 
\bibitem{WEISEMANN1}S.~Wesemann, H.~Schlesinger, A.~Pascht, and O.~Blume, ``Measurement and characterization of the temporal behavior of fixed massive MIMO links," in \emph{21th International ITG Workshop on Smart Antennas (WSA)}, Mar. 2017, pp. 135-142. 
\bibitem{ICA} A. Hyvarinen, J. Karhunen, and E. Oja,  \emph{Independent Component Analysis}. John Wiley \& Sons, 2001.
\bibitem{b8} S. M. Kay, \emph{Fundamentals of Statistical Processing Vol. I: Estimation Theory}, Englewood Cliffs, NJ, USA: Prentice-Hall, 1993.
\bibitem{matrix}R. A. Horn and C. R. Johnson, \emph{Matrix Analysis}, Cambridge University Press, 1985. 
\bibitem{b10} M. Loog, "On Distributional assumptions and whitened cosine similarities," \emph{IEEE Trans. Pattern Analysis and Machine Intelligence}, vol. 30, no. 6, pp. 1114-1115, Jun. 2008.
\end{thebibliography}
\end{document}